\documentclass[12pt,preprint]{aastex}

\usepackage{color}
\shorttitle{GWs from IMBHs in YCs}
\shortauthors{Mapelli et al.}

\begin{document}


\title{Gravitational waves from intermediate-mass black holes in young clusters}


\author{M. Mapelli\altaffilmark{1,2}, C. Huwyler\altaffilmark{1}, L. Mayer\altaffilmark{1}, Ph. Jetzer\altaffilmark{1}, A. Vecchio\altaffilmark{3}}
%\author{C. Huwyler\altaffilmark{1}, M. Mapelli\altaffilmark{1}, L. Mayer\altaffilmark{1}, P. Jetzer\altaffilmark{1}}
\altaffiltext{1}{Institute for Theoretical Physics, University of Zurich, Winterthurerstrasse 190, CH 8057 Zurich, Switzerland}
\altaffiltext{2}{Dipartimento di Fisica G. Occhialini, Universit\`a degli Studi di Milano-Bicocca, Piazza della Scienza 3, I-20126, Milano, Italy}
\altaffiltext{3}{School of Physics and Astronomy, University of Birmingham, Edgbaston, Birmingham B15 2TT, UK}



\begin{abstract}
Massive young clusters (YCs) are expected to host intermediate-mass black holes (IMBHs) born via runaway collapse. 
These IMBHs are likely in binaries and can undergo mergers with other compact objects, such as stellar mass black holes (BHs) and neutron stars (NSs). 
We derive the frequency of such mergers starting from information available in the Local Universe; in particular, we assume that a fraction $\sim{}0.75$ of all the YCs more massive than $10^5$ M$_{\odot{}}$ might host one IMBH, as suggested by a statistical analysis of the properties of YCs in the Milky Way and in the Antennae. 
 Mergers of IMBH$-$NS and IMBH$-$BH binaries are sources of gravitational waves (GWs), which might allow us to reveal the presence of IMBHs.
We thus examine their 
detectability by current and future GW 
observatories, both ground- and space-based. In particular, as representative of different classes of instruments we consider Initial and Advanced LIGO, the Einstein gravitational-wave Telescope (ET) and the Laser Interferometer Space Antenna (LISA).
We find that IMBH mergers are unlikely to be detected with instruments operating at the current sensitivity (Initial LIGO). 
 LISA detections are disfavored by the mass range of IMBH$-$NS and IMBH$-$BH binaries: less than one event per year is expected to be observed by such instrument.
Advanced LIGO is expected to 
observe a few merger events involving IMBH binaries in a 1-year long observation. Advanced LIGO is particularly suited for mergers of relatively light IMBHs ($\sim{}10^2\,{}{\rm M}_\odot{}$) with stellar mass BHs. 
The number of mergers detectable with ET is much larger: 
tens (hundreds) of IMBH$-$NS (IMBH$-$BH) mergers might be observed per year, according to the runaway collapse scenario for the formation of IMBHs.  
We note that our results are affected
by large uncertainties, produced by poor observational constraints on many
of the physical processes involved in this study, such as the evolution of
the YC density with redshift.

\end{abstract}


%\keywords{gravitational waves --- black hole physics  --- stellar dynamics --- galaxies: star clusters}
\keywords{gravitational waves --- black hole physics  --- galaxies: star clusters: general  --- (stars:) binaries: general --- stars: kinematics and dynamics}



\section{Introduction}
So far, there are no definitive observational proofs of the existence of intermediate-mass black holes (IMBHs), i.e. of black holes (BHs) with a mass ranging from $\sim{}10^2$ to $\sim{}10^5\,{}{\rm M}_\odot{}$ (see van der Marel 2004 for a review). However, the formation of IMBHs is predicted by various theoretical scenarios.  In the early Universe, IMBHs might form via the direct collapse of metal-free stars (Heger et al. 2003) or via the collapse of gaseous disks in the center of pre-galactic halos (Begelman, Volonteri \&{} Rees 2006). IMBHs might form even in more recent epochs, via the direct collapse  of metal-poor stars (Mapelli, Colpi \&{} Zampieri 2009; Zampieri \&{} Roberts 2009; Mapelli et al. 2010), the repeated mergers of stellar mass BHs in globular clusters (GCs, Miller \&{} Hamilton 2002) and the runaway growth of IMBHs (Portegies Zwart \& McMillan 2002, hereafter PZM02) in massive ($\ge{}10^4\,{}{\rm M}_\odot{}$) young ($\lesssim{}3\times{}10^7$ yr) clusters (YCs). In particular, the runaway growth scenario predicts that star clusters with initial half-mass relaxation times $t_{\rm h}\lesssim{}25$ Myr are dominated, during the core collapse, by star collisions, which build up one or even two (G\"urkan, Fregeau \&{} Rasio 2006) very massive objects, which likely evolve into IMBHs (PZM02). Thus, YCs are among the best environments where we can search for IMBHs. 

However, the observational features of IMBHs are  difficult to pinpoint. There are hints (Strohmayer \&{} Mushotzky 2003; Kaaret, Ward \&{} Zezas 2004; Strohmayer et al. 2007) that IMBHs might power some of the brightest ultra-luminous X-ray sources (ULXs, i.e. those sources with a X-ray luminosity higher than the one predicted for a $\sim{}10\,{}{\rm M}_\odot{}$ BH). Other studies point out that the presence of IMBHs may be inferred from anomalies in the acceleration of millisecond pulsars (Colpi, Possenti \&{} Gualandris 2002; Colpi, Mapelli \&{} Possenti 2003) or from the observation of high-velocity stars in GCs (Mapelli et al. 2005). Furthermore, the combination of kinematic and photometric data with dynamical models suggests the presence of IMBHs in the GC G1 (Gebhardt, Rich \&{} Ho 2002, 2005) in the Andromeda galaxy and in the GC M15 (Gerssen et al. 2002; van den Bosch et al. 2006) in the Milky Way (MW). 

Finally, gravitational waves (GWs) will provide a powerful tool to detect IMBHs. In fact, IMBHs lying in the center of dense clusters are expected to form binaries with stars or stellar mass BHs (e.g. Sigurdsson \&{} Hernquist 1993). Such binaries harden (i.e. progressively reduce their orbital separation) due to gravitational encounters (Colpi et al. 2003 and references therein). When the  orbital separation is sufficiently small and if the companion of the IMBH is a compact object, such as a stellar mass BH or a neutron star (NS), the binary enters a regime in which the orbital evolution is driven by radiation reaction induced by GW emission.
This further reduces the orbital separation, until the system merges. GWs emitted by binaries hosting  IMBHs are mostly in the frequency range accessible to current or forthcoming detectors, such as the Laser Interferometer Gravitational-Wave Observatory (LIGO; Barish \& Weiss 1999; Abbott et al. 2009) and Virgo (Acernese et al. 2004) in initial and advanced configuration (also known as first and second generation laser interferometers). GWs from IMBH mergers also fall in the frequency range of future GW observatories, such as the Laser Interferometer Space Antenna (LISA; Bender et al. 1998) and third generation ground-based instruments, for which we adopt, as an example, the concept of the Einstein gravitational-wave Telescope (ET; Freise et al 2008, Hild et al. 2008). In particular, Miller (2002), considering IMBHs in GCs, predicts that tens of sources connected with Galactic IMBHs will be detectable with LISA in a 5 yr integration. More recent studies investigate the emission of GWs from IMBH$-$IMBH binaries (Fregeau et al. 2006), whose existence is predicted by simulations (G\"urkan et al. 2006). Massive stellar mass BHs ($15-20\,{}{\rm M}_\odot{}$) are also found to be important sources of GWs, detectable by LIGO and by Advanced LIGO (O'Leary, O'Shaughnessy \&{} Rasio 2007).
 Inspirals of a NS or of a stellar mass BH into an IMBH in GCs are detectable by Advanced LIGO, with rates up to tens per year (Brown et  al. 2007; Mandel et al. 2008).  Finally, ET might be able to observe GWs from IMBHs produced by seed BHs at high redshift
(Sesana et al. 2009; Gair et al. 2009a, 2009b).
 
In this Paper we extend the previous studies by Miller (2002) and by Mandel et al. (2008) to IMBHs hosted in YCs, making predictions for the detection rate of GWs from IMBHs in YCs. YCs are an ideal environment in which GWs from IMBHs can be studied, as YCs are a key place for the formation of IMBHs and because the density of YCs is expected to rapidly increase with redshift (Hopkins \&{} Beacom 2006). For this reason, we pay particular attention to the integration with redshift of the detection rate. We consider, as representative of the science capability of first, second and third generation ground-based instruments Initial LIGO, Advanced LIGO and ET, respectively, and for space-borne instruments LISA. We note that space-based instruments with peak sensitivity in the frequency region $\sim 0.1$ Hz that fills the gap between LISA and ground based laser interferometers would be ideal for studying IMBHs, due to the emission frequency of systems in the mass range $\sim 10^2 - 10^3 \,{\rm M}_\odot$. Such instruments, such as DECIGO (Kawamura et al. 2006), ALIA (Bender et al. 2005)  and the Big-Bang-Observer (Phinney et al. 2003) are being investigated, but their timescale is currently very uncertain and for this reason we will not consider them in this Paper.

\section{Results}
In this Section, we will focus on the properties of IMBHs in YCs and on their role as sources of GWs. 
In particular, we will consider two galaxies: the MW, for which we have the best observational data about YCs, and the Antennae, which host a large population of YCs. 
These two galaxies can also be considered as prototypical of, respectively, a bright disk galaxy with a typical star formation (SF) rate for the current cosmic epoch, 
and a merging system involving two massive disk galaxies undergoing enhanced SF, a configuration that is expected to be increasingly common at higher redshift.
We will subsequently generalize our results and make predictions for the detection rate of GWs by LIGO, Advanced LIGO, LISA and ET.

\subsection{IMBH mergers in the MW}
First, we derive an estimate of the number of IMBHs which may exist in the YCs of the MW.
As the YCs reside in the high-density regions of our Galaxy,
their detection and observation are made difficult by Galactic absorption. Thus, only data for $\sim{}10$ massive ($\ge{}10^4\,{}{\rm M}_\odot{}$) Galactic YCs are available (Davies et al. 2007 and references therein). However, Gvaramadze, Gualandris \&{} Portegies Zwart (2008) estimate that the total number of massive Galactic YCs is $\sim{}70-100$, much higher than the number of the detected ones. The runaway growth of IMBHs is possible only in those YCs with half-mass relaxation time $t_{\rm h}\lesssim{}25$ Myr (PZM02). Extrapolating from the properties of the YCs listed in table 1 of Portegies Zwart et al. (2002), we find that 
a fraction $f_{\rm tot}\sim{}0.5-1.0$ of the Galactic massive YCs have $t_{\rm h}\lesssim{}25$ Myr and thus may host IMBHs. Therefore, $\sim{}30-100$ IMBHs may exist at present in the YCs of the MW. In the following sections, we adopt the average value $f_{\rm tot}=0.75$ as a fiducial value.

IMBHs born via runaway collapse in the center of dense YCs likely form binaries with other stars or stellar mass BHs (see Colpi et al. 2003 and references therein). Such binaries harden due to three-body (or even four-body) interactions at a rate (Sigurdsson \&{} Phinney 1993; Colpi et al. 2003)
\begin{equation}\label{eq:eq3body}
\nu_{3b}\sim{} 2\,{}\pi{}\,{}G\,{}m_{\rm BH}\,{}n_{\rm c}\,{}a\,{}\sigma{}_{\rm c}^{-1}
 \sim{} 10^{-6}\,{}{\rm yr}^{-1}\,{}\left(\frac{m_{\rm BH}}{500\,{}{\rm M}_\odot{}}\right)\,{}\left(\frac{n_{\rm c}}{5\times{}10^5\,{}{\rm pc}^{-3}}\right)\,{}\left(\frac{a}{0.4\,{}{\rm A.U.}}\right)\,{}\left(\frac{20\,{}{\rm km}\,{}{\rm s}^{-1}}{\sigma{}_{\rm c}}\right),
\end{equation}
where $G$ is the gravitational constant, $m_{\rm BH}$ is the mass of the IMBH, $n_{\rm c}$ is the core density of the YC, $a$ is the orbital separation of the binary and $\sigma{}_{\rm c}$ is the velocity dispersion in the core of the YC. In equation~(\ref{eq:eq3body}) we take as fiducial values for the MW $n_{\rm c}=5\times{}10^5$ stars pc$^{-3}$ and $\sigma{}_{\rm c}=20$ km s$^{-1}$, which are the average values for the YCs listed in table 1 of Portegies Zwart et al. (2002). The masses of IMBHs formed by runaway collapse ($m_{\rm BH}$) likely range from  $10^2$ to $10^3\,{}{\rm M}_\odot{}$ (PZM02; G\"urkan et al. 2006). 
Finally, a typical initial orbital separation of $\sim{}0.4$ A.U. is assumed, according to Devecchi et al. (2007)\footnote{The fiducial value $a=0.4$ A.U., adopted in our calculations, is the typical orbital separation of an IMBH$-$BH (or IMBH$-$NS) binary which is already hard (i.e. whose binding energy is larger than the average kinetic energy of a star in the cluster, see e.g. Heggie 1975), but  which is still in the regime where the hardening due to gravitational encounters is more important than that due to gravitational wave emission. For more details about the derivation of $a$, see Table~1, equations $1-3$ and the Appendix~A of Devecchi et al. (2007). We note that a more realistic approach should consider a distribution of values for $a$, depending on $m_{\rm BH}$, on the mass of the companion, on $n_{\rm c}$, on $\sigma{}_{\rm c}$, and on the history of previous gravitational encounters. Here, we simply adopt the peak of the distribution of $a$ for m$_{\rm BH}\sim{}300$ M$_\odot{}$, as derived by Devecchi et al. (2007), in order to provide an approximate estimate of $\nu_{3b}$.}.

 Typically, a merger between the IMBH and its companion occurs after a few hundred three-body interactions (Miller 2002). Since it is hard to give a more accurate expression for the merger rate, in the following we will assume that  the merger rate for a single IMBH binary is $\nu{}_{\rm mrg}=10^{-2}\,{}\nu_{3b}$ (Miller 2002), neglecting its possible dependence on various quantities, such as m$_{\rm BH}$. Thus, the merger rate for a single IMBH binary in the Milky Way is $\nu{}_{\rm mrg}\sim{}10^{-8}\,{}(m_{\rm BH}/500\,{}{\rm M}_\odot{})\,{}(n_{\rm c}/5\times{}10^5\,{}{\rm pc}^{-3})\,{}(a/0.4\,{}{\rm A.U.})\,{}(20\,{}{\rm km}\,{}{\rm s}^{-1}/\sigma{}_{\rm c})$ yr$^{-1}$. We expect a Galactic total merger rate $\nu{}_{\rm mrg, tot}\approx{} 10^{-6}\,{}{\rm yr}^{-1}$, considering all the $\sim{}100$ IMBHs hosted in the YCs of the MW.

\subsection{IMBH mergers in the Antennae}

The Antennae are a well-studied nearby galaxy pair which is undergoing merger. As the merger  triggers the SF, the Antennae are richer in YCs than the MW. In fact, more than 1000 YC candidates have been observed in the Antennae (Mengel et al. 2005). The expected number of massive YCs in the Antennae can be derived with the same calculation used by Gvaramadze et al. (2008) for the MW. In particular, considering a SF rate (SFR) of $7.1\,{}{\rm M}_\odot{}$ yr$^{-1}$ for the Antennae (Grimm, Gilfanov \&{} Sunyaev 2003), assuming an upper limit of $\sim{}3\times{}10^7$ yr for the age of the YCs and using a power-law cluster initial mass function (MF) with slope 2 (Gvaramadze et al. 2008), we find that $\sim{}2100$ massive YCs are expected to exist in the Antennae at present. Assuming that the half-mass relaxation time for the YCs in the Antennae is similar to that of the YCs in the MW\footnote{This assumption is reasonable, as various properties of the YCs in the Antennae (e.g. velocity dispersion, total cluster mass, etc.) are similar to the ones of the YCs in the MW (Mengel et al. 2002; de Grijs et al. 2005).}, we expect that $\sim{}1000-2000$ IMBHs exist in the Antennae. Thus, the expected merger rate for IMBHs in the Antennae is $\nu{}_{\rm mrg, tot}\approx{} 10^{-5}\,{}{\rm yr}^{-1}$, i.e. a factor of $\gtrsim{}10$ higher than for the MW.

\subsection{GWs from IMBHs in the local Universe}
From the calculations reported in the previous Sections, we can predict a detection rate of GWs emitted by merging IMBHs in the local Universe, expected for LISA, LIGO, Advanced LIGO and ET.  In particular, we consider all the mergers involving IMBH$-$BH and IMBH$-$NS binaries on the past light cone, within a sphere of fixed radius, derived from the detection range of a given instrument, through the expression of the
maximum observable redshift $z_{\rm max}$, see the description below. 
Under the assumption that all the IMBHs have the same mass, such detection rate can be expressed as (Miller 2002):

\begin{equation}\label{eq:GW}
R=4\,{}\pi{}\,{}\left(\frac{c}{H_0}\right)^3\,{}f_{\rm tot}\,{}\nu{}_{\rm mrg}(m_{\rm BH})\,{}\int\limits^{z_{\rm max}(m_{\rm BH},\,{}m_{\rm co})}_{0}\left(\int\limits^{z}_{0}\frac{{\rm d}\tilde{z}}{E(\tilde{z})}\right)^2\,{}\frac{n_{\rm YC}(z, m_{\rm BH})}{E(z)}\,{}\frac{{\rm d}t_{\rm e}}{{\rm d}{t_{\rm o}}}\,{}{\rm d}z
\end{equation}
where $c$ is the light speed, $H_0$ ($=72$ km$/$s Mpc$^{-1}$, Spergel et al. 2007) the Hubble constant, $f_{\rm tot}\sim{}0.75$ is the fraction of massive YCs hosting an IMBH (see Section 2.1), $\nu_{\rm mrg}(m_{\rm BH})$ is the merger rate per IMBH (see Section 2.1), $E(z)=\left[(1+z)^3\Omega{}_{\rm M}+\Omega{}_\Lambda\right]^{1/2}$ (where $\Omega{}_{\rm M}=0.27$ and $\Omega{}_{\rm \Lambda}=0.73$, Spergel et al. 2007). The factor $\frac{{\rm d}t_{\rm e}}{{\rm d}{t_{\rm o}}}=(1+z)^{-1}$ accounts for the difference between the time in the rest frame of the source ($t_{\rm e}$) and of the observer ($t_{\rm o}$).

In equation~(\ref{eq:GW}), $n_{\rm YC}(z, m_{\rm BH})$ is the comoving number density of YCs  which are sufficiently massive to host an IMBH. Assuming that a constant fraction of stars ($f_{\rm SFC}$) forms in clusters, $n_{\rm YC}(z, m_{\rm BH})$ can be approximately estimated as:
\begin{eqnarray}\label{eq:eqYC}
n_{\rm YC}(z,m_{\rm BH})=4.5\times{}10^{-2}\,{}{\rm Mpc}^{-3}\,{}\left(\frac{\dot{\rho{}}_{\ast{}}(z)}{1.5\times{}10^{-2}\,{}{\rm M}_\odot{}{\rm yr}^{-1}{\rm Mpc}^{-3}}\right)\,{} \nonumber{}\\
\quad{}\quad{}\quad{}\times{}\left(\frac{t_{\rm max}}{10^9{\rm yr}}\right)\,{}\left(\frac{f_{\rm surv}}{10^{-2}}\right)\,{}\left(\frac{\langle{}m_{\rm YC}\rangle{}}{260\,{}{\rm M}_\odot{}}\right)^{-1}\,{}\left(\frac{f_{\rm SFC}}{0.8}\right)\,{}\left(\frac{f_{\rm YC}(m_{\rm BH})}{10^{-4}}\right),
\end{eqnarray}
where $t_{\rm max}$ is the maximum lifetime of a YC: we adopt $t_{\rm max}=10^9$ yr, as we account for the fact that the IMBH$-$BH and IMBH$-$NS mergers occur $\sim{}2\times{}10^8$ yr after the formation of the central binary. We must also include a correction $f_{\rm surv}$ which represents the fraction of YCs which avoid disruption and survive up to $t_{\rm max}=10^9$ yr. $f_{\rm surv}$ is quite uncertain; we adopt a conservative value $f_{\rm surv}=10^{-2}$ from the literature (Lada \&{} Lada 2003; Fall et al. 2005). For a discussion about the possible mechanisms of cluster infant mortality, see, e.g., Gieles (2010) and references therein.

 $\dot{\rho{}}_{\ast{}}(z)=10^{\left[\beta{}\,{}\log_{10}{(1+z)}+\alpha{}\right]}$ is the comoving density of SFR (where $\alpha{}=-1.82$ and $\beta{}=3.28$ for $z\le{}1.04$, and $\alpha{}=-0.724$ and $\beta{}=-0.26$ for $1.04\le{}z\le{}4.48$,  Hopkins \&{} Beacom 2006). These values of $\dot{\rho{}}_{\ast{}}(z)$ are based on the most updated available data\footnote{We note that the assumption that the YC density is proportional to the SFR is quite inconsistent with the fact that we assume a lifetime $t_{\rm max}=10^9$ yr for the YCs, likely overestimating high-redshift clusters ($z\sim{}4$) by $\lesssim{}30$ per cent. On the other hand, it is quite hard to refine this model (given our poor knowledge of high-redshift YCs). Furthermore, values of the redshift higher than 1 are important only for ET (see Fig.\ref{fig:fig1}). Finally, we adopted a conservative value of $f_{\rm surv}$ ($=10^{-2}$), in order to avoid overestimating $n_{\rm YC}$.}.
$f_{\rm SFC}$ is the fraction of SF which occurs in clusters ($f_{\rm SFC}=0.7-0.9$, Lada \&{} Lada 2003). We assume that $f_{\rm SFC}$ is constant with redshift, although this cannot be proved on the basis of currently available data.

$\langle{}m_{\rm YC}\rangle{}\sim{}260\,{}{\rm M}_\odot{}$ is the average mass of a YC, and has been derived assuming that the masses of the YCs range from $m_{\rm YC,\,{}min}\sim{}20\,{}{\rm M}_\odot{}$ to $m_{\rm YC,\,{}max}\sim{}10^7\,{}{\rm M}_\odot{}$ and are distributed according to a  MF $\frac{{\rm d}N}{{\rm d}m}\propto{}m^{-2}$ (Lada \&{} Lada 2003).
Finally,  $f_{\rm YC}(m_{\rm BH})$ is the fraction of YCs which are massive enough to host IMBHs: 
\begin{equation}
f_{\rm YC}(m_{\rm BH})=\left(\int_{m_{\rm YC,\,{}min}}^{m_{\rm YC,\,{}max}} {\frac{{\rm d}N}{{\rm d}m}\,{}{\rm d}m}\right)^{-1}\,{}\int_{10^3\,{}m_{\rm BH}}^{m_{\rm YC,\,{}max}}{\frac{{\rm d}N}{{\rm d}m}\,{}{\rm d}m}\,;
\end{equation}
$f_{\rm YC}$ depends on $m_{\rm BH}$, as the simulations show that the IMBH mass scales with the cluster mass ($m_{\rm BH}\sim{}10^{-3}\,{}m_{\rm YC}$, PZM02). In the following calculations, we assume that $m_{\rm YC,\,{}min}=20\,{}{\rm M}_\odot{}$, that $m_{\rm YC,\,{}max}=10^7\,{}{\rm M}_\odot{}$ and that $\frac{{\rm d}N}{{\rm d}m}\propto{}m^{-2}$ (Lada \&{} Lada 2003). Under these assumptions, $f_{\rm YC}(m_{\rm BH})$ goes from $\sim{}2\times{}10^{-4}$ (for an IMBH mass $m_{\rm BH}=100\,{}{\rm M}_\odot{}$) to $\sim{}1.8\times{}10^{-5}$ (for an IMBH mass $m_{\rm BH}=1000\,{}{\rm M}_\odot{}$).

In equation~(\ref{eq:GW}), $z_{\rm max}(m_{\rm BH},\,{}m_{\rm co})$ is the maximum redshift at which an event can be detected with a sky-location and orientation averaged signal-to-noise ratio $\langle{\mathrm SNR}\rangle \ge{}10$ by a single interferometer. In observations with a network of instruments, the signal-to-noise ratio scales as the square root of the number of instruments, and in this respect the results presented here should be considered as conservative.
The maximum redshift depends on the mass of the IMBH $m_{\rm BH}$ and on the mass of the companion which merges with the IMBH ($m_{\rm co}$), as well as on the sensitivity of the instrument. 
The expressions adopted to calculate $z_{\rm max}(m_{\rm BH},\,{}m_{\rm co})$ (for LIGO, Advanced LIGO, LISA and ET) are summarized in Appendix~A. 
As a model for the gravitational waveform, we consider the analytical  
phenomenological inspiral-merger-ringdown waveform approximant for non- 
spinning BHs in circular orbits derived by Ajith et at. (2008a). This  
waveform model together with the Effective-One-Body-Numerical- 
Relativity (EOBNR) waveform family (Buonanno et al 2007) provides a  
prescription for the signal from the whole coalescence of binary  
systems. These waveform approximants have both been calibrated on full  
numerical relativity simulations for binaries for approximately equal  
mass systems, and yield signal-to-noise ratios that are consistent  
within $\approx 25\%$. However, in this paper we will apply the  
waveforms on a much larger mass-ratio regime that extends up to $\sim  
10^{-2}$. For this mass-ratio we still do not have reliable inspiral- 
merger-ring-down waveforms, and the phenomenological and EOBNR  
approximants yield significantly different signal-to-noise ratios,  
that differ by a factor $\approx$ 2$-$10 depending on the mass ratio,  
with the phenomenological waveforms producing the larger signal-to-noise ratio, see \emph{e.g.} Fig. 2 of Gair et al (2009b). As the detection rate scales as the cube of the
maximum distance at which a source can be detected, the amplitude
differences lead to rate uncertainties of a factor 10 or larger. 
%\footnote{\bf We remind that, since the amplitudes calculated with phenomenological approximants differ by a factor of 2$-$10 from those obtained with the EOBNR  approximants, the differences between the derived detection rates are expected to be even larger.}. 
 Furthermore  
if the BHs are (rapidly) spinning, the signal-to-noise ratio at which  
a source is observed can be significantly (by a factor of $\approx 2$,  
see \emph{e.g.} Ajith et al. 2009) affected. However, the lack of  
astrophysical predictions for the likely spins of IMBH binaries and  
the lack of full coalescence waveform approximants for generic spin  
magnitudes and geometries (see however Ajith et al. 2009; Pan et al.  
2009) prevent us from considering this possibly important physical  
effect.
The resulting values of $z_{\rm max}(m_{\rm BH},\,{}m_{\rm co})$ are shown in the upper panels of Fig.~\ref{fig:fig1}. In this Fig. we consider mergers of the IMBHs with two different compact objects, i.e. NSs (left-hand panel) and stellar mass BHs (right-hand panel)\footnote{In our calculations we assume a mass of $m_{\rm co}=1.4$ and $m_{\rm co}=10\,{}{\rm M}_\odot{}$ for the NSs and the stellar mass BHs, respectively. 
Furthermore, we  assume that all the IMBHs formed in YCs merge with a compact object. This assumption is justified by the strong mass segregation occurring during the runaway collapse (PZM02). The strong mass segregation and the concentration of the most massive stellar objects at the center of the cluster are confirmed by the observed mass distribution in the core of some Galactic globular clusters (where the mass distribution within the core is $\frac{dN}{dm}\propto{}m^{\alpha{}}$, with $\alpha{}=3-8$, see e.g. Prior et al. 1986 and  Monkman et al. 2006). These evidences, combined with the fact that hard binaries (i.e. binaries with binding energy larger than the average kinetic energy of a cluster star) tend to exchange, during three-body encounters, with the most massive possible companion (see e.g. Sigurdsson \&{} Phinney 1995), suggest that most of IMBHs formed in YCs merge with compact objects.}.
It is evident that there is a large difference between ET and the other interferometers. In fact, while for ET $z_{\rm max}(m_{\rm BH},\,{}m_{\rm co})$ is always higher than 0.5 and 1 in the case of NSs and stellar mass BHs, respectively, for Advanced LIGO $z_{\rm max}(m_{\rm BH},\,{}m_{\rm co})<0.2$ and $<0.5$ in the case of  NSs and stellar mass BHs, respectively. For the other interferometers $z_{\rm max}(m_{\rm BH},\,{}m_{\rm co})$ is even smaller.

 The detection rates $R$ derived from equation~(\ref{eq:GW}) for events detectable with Advanced LIGO, LISA and ET are shown in the bottom panels of Fig.~\ref{fig:fig1}. In particular, the detection rates $R$ for mergers involving IMBH$-$NS and IMBH$-$BH binaries are shown in the left-hand panel and in the right-hand panel, respectively. The different lines show the values of $R$ for different detectors, as a function of the IMBH mass, ranging from $10^2$ to $10^3\,{}{\rm M}_\odot$. The results obtained for LIGO are not shown in  Fig.~\ref{fig:fig1}, as they are orders of magnitude lower than those obtained for the other interferometers. In particular, the rate of IMBH$-$NS (IMBH$-$BH) mergers detectable with LIGO is lower than $5\times{}10^{-5}$ ($1\times{}10^{-3}$) per year, even in the most favorable case (corresponding to $m_{\rm BH}=100\,{}{\rm M}_\odot{}$).

 Fig.~\ref{fig:fig1}  shows that ET (dotted line, red on the web) is expected to detect a large number of events per year: $R>10$ yr$^{-1}$ and $R>60$ yr$^{-1}$ in the case of IMBH$-$NS and IMBH$-$BH mergers, respectively. Thus, ET will be essential to assess the existence of IMBHs in the Local Universe, and to distinguish among different theories of IMBH formation. Advanced LIGO (solid black line) is expected to detect a few mergers involving IMBH$-$BH binaries per year and less than one merger involving IMBH$-$NS binaries per year. Advanced LIGO is particularly 
suitable for observations of low-mass IMBHs ($\lesssim{}300\,{}{\rm M}_\odot{}$) merging with stellar mass BHs. In the case of $m_{\rm BH}=100\,{}{\rm M}_\odot{}$, Advanced LIGO is expected to detect up to $\sim{}5$ IMBH$-$BH mergers per year.

On the other hand, LISA (dashed line, green on the web) is more efficient in detecting massive ($\gtrsim{}500\,{}{\rm M}_\odot{}$) IMBHs merging with stellar mass BHs, due to its sensitivity to frequencies lower than 1 Hz. However, LISA is not particularly suited for detecting GWs from IMBHs, as less than one event per year is expected to be detected, 
even in the most optimistic case (i.e. for $m_{\rm BH}=10^3\,{}{\rm M}_\odot{}$). In the case of IMBH$-$NS mergers, LISA is very unlikely to observe any such system during the mission lifetime (nominally set to 5 years), even for $m_{\rm BH}=10^3\,{}{\rm M}_\odot{}$; the detection rate is in fact $\lesssim{}10^{-4}\,\mathrm{yr}^{-1}$.

In equation~(\ref{eq:GW}) we assume that all the IMBHs have the same mass.
However, this is 
 an over-simplification, as IMBH masses will be distributed according to a certain MF.
 Accounting for such MF, equation~(\ref{eq:GW}) becomes:
\begin{eqnarray}\label{eq:GW2}
R=4\,{}\pi{}\,{}\left(\frac{c}{H_0}\right)^3\,{}f_{\rm tot}\,{}\left(\,{}\int\limits^{m_2}_{m_1}\,{}\frac{{\rm d}N}{{\rm d}m_{\rm BH}}\,{}{\rm d}m_{\rm BH}\right)^{-1}\times{}\quad{}\quad{}\quad{}\quad{}\quad{}\quad{}\quad{}\quad{}\quad{}\quad{}\quad{}\quad{}\quad{}\quad{}\nonumber  \\
\int\limits^{m_2}_{m_1}\,{}\nu{}_{\rm mrg}(m_{\rm BH})\,{}\left[\int\limits^{z_{\rm max}(m_{\rm BH},\,{}m_{\rm co})}_{0}\left(\int\limits^{z}_{0}\frac{{\rm d}\tilde{z}}{E(\tilde{z})}\right)^2\,{}\frac{n_{\rm YC}(z,m_{\rm BH})}{(1+z)\,{}E(z)}\,{}{\rm d}z\right]\,{}\frac{{\rm d}N}{{\rm d}m_{\rm BH}}\,{}{\rm d}m_{\rm BH},
\end{eqnarray}
where $\frac{{\rm d}N}{{\rm d}m_{\rm BH}}$ is the MF of the IMBHs, whereas $m_1$ and $m_2$ are the minimum and the maximum IMBH mass, respectively.

The MF of IMBHs is unknown. However, as we already mentioned for the estimate of $f_{\rm YC}(m_{\rm BH})$,  PZM02 show that there is a correlation between the mass of the IMBH and the mass of the host YC ($m_{\rm YC}$), and in particular that the mass of the IMBH is $m_{\rm BH}\sim{}10^{-3}m_{\rm YC}$. Since the YCs in the MW have a MF $\frac{{\rm d}N}{{\rm d}m}\propto{}m^{-2}$ (Lada \&{} Lada 2003), we can assume that the IMBHs have the same MF. Adopting this MF and assuming $m_1=100\,{}{\rm M}_\odot{}$ and $m_2=1000\,{}{\rm M}_\odot{}$, we find $\langle{}m_{\rm BH}\rangle{}=256\,{}{\rm M}_\odot{}$.

Integrating equation~(\ref{eq:GW2}) for these values of the MF and of $m_1$ and $m_2$, we obtain the detection rates $R$ shown in Table~1. In Fig.~\ref{fig:fig1} we also plot the results of equation~(\ref{eq:GW2}), using a filled square (red on the web), a filled black circle  and a filled triangle (green on the web) for ET, Advanced LIGO and LISA, respectively. These points are drawn, for simplicity, in correspondence of $m_{\rm BH}=256\,{}{\rm M}_\odot$, but they have been obtained integrating equation~(\ref{eq:GW2}) over a Lada \&{} Lada (2003) MF, with an average IMBH mass $\langle{}m_{\rm BH}\rangle{}=256\,{}{\rm M}_\odot{}$. We note that the results obtained from equation~(\ref{eq:GW2}), under these assumptions, are very similar to those obtained from equation~(\ref{eq:GW}), assuming that all the IMBHs have mass $m_{\rm BH}=256\,{}{\rm M}_\odot{}$.

  Fig.~\ref{fig:fig1} and Table~1  show that, if the IMBHs are distributed according to a Lada \&{} Lada (2003) MF, ET is expected to detect a large number of events per year, involving both IMBH$-$NS ($R\sim{}200$ yr$^{-1}$) and IMBH$-$BH ($R\sim{}400$ yr$^{-1}$) binaries. Advanced LIGO is expected to detect $\sim{}4$ mergers involving IMBH$-$BH binaries and $\sim{}0.2$ mergers involving IMBH$-$NS binaries per year. 
LISA is not suited for detecting IMBHs distributed according to the Lada \&{} Lada (2003) MF, as such MF favors relatively `light' IMBHs.

In Appendix~B we derive an alternative calculation of $R$, based on equation~(14) of Gair et al. (2009b). In Fig.~\ref{fig:fig1}, the results of the alternative calculation reported in Appendix~B are shown as open points: an open square in the case of ET, an open circle for Advanced LIGO, and an open triangle for LISA. The results of the alternative calculation are similar (within a factor of 5) to those of  equation~(\ref{eq:GW2}). Given the large uncertainties in both calculations, the results are consistent between them. Furthermore, a more simplified,  order-of-magnitude derivation of the detection rate $R$ is provided in Appendix~C. We stress that the calculation in Appendix~C is much more approximated than the one presented in this Section and the one reported in Appendix~B, but it is a useful tool, in order to estimate the order of magnitude of $R$ with `back of the envelope' calculations.

 Our results are consistent with those indicated by previous studies.
In particular, in the case of LISA, our estimates of the detection rate $R$ are approximately one order of magnitude higher than those obtained in Will (2004), which adopts the noise curves by Larson et al. (2000), and a factor of $\lesssim{}5$ lower than those derived by Miller (2002), based on the noise curves in Flanagan \&{} Hughes (1998a, 1998b).
The results obtained for Advanced LIGO are consistent with the recent study by Mandel et al. (2008). 
In the case of ET, as we discuss in Appendix~C, the results of our approximate calculation in equation~(\ref{eq:approx}) are very similar (within a factor of two) to those reported in Table~1 of Gair et al. (2009b). The results derived from the more refined equation~(\ref{eq:GW2}) and from the alternative model in equation~(\ref{eq:gair4}) are a factor of $4-10$ larger than those reported in Gair et al. (2009b), mainly due to the fact that we integrate over the IMBH mass function and that we account for the dependence of n$_{\rm YC}$ on the redshift.

In conclusion, GWs from IMBHs are hardly detectable with LIGO and LISA. Instead, Advanced LIGO has chances of detecting GWs from IMBHs in clusters. ET is expected to observe hundreds of IMBHs formed via the runaway collapse mechanism.

\section{Summary}
In this Paper we study the occurrence of mergers between IMBHs and compact objects (NSs and stellar mass BHs) in YCs. 
These are found to be important sources of GWs.  Our study shows that GWs from IMBHs 
are unlikely to be detected with first generation instruments, such as Initial LIGO, and even with LISA. Advanced LIGO 
 offers instead the ability of observing these sources: a few merger events are expected to be detected by Advanced LIGO in 1-year integration. ET is far the best interferometer projected up to now to study GWs from IMBHs: $\sim{}10$ to $\sim{}300$ ($\sim{}60$ to $\sim{}600$) merger events of IMBH$-$NS (IMBH$-$BH) binaries might be detected in 1-year integration, according to the scenario of runaway collapse for IMBH formation. Thus, ET will be a powerful tool to check the runaway collapse and the other scenarios for IMBH formation.
However, our estimates are affected by large uncertainties (more than a factor of 10), because of our insufficient knowledge 
of masses and density of IMBHs in the local Universe. In particular, the fraction of IMBHs per cluster $f_{\rm tot}$ is highly uncertain, as it is based only on the properties of the $\sim{}5$ Galactic YCs for which enough data are available. Furthermore, $f_{\rm tot}=0.75$ adopted in 
this Paper must be considered as an upper limit, as we are assuming that all the YCs which can undergo runaway collapse host an IMBH, and that all the mass involved in the collapse ends up into the IMBH. There 
are no evidences that the runaway collapse necessarily leads to the formation of an IMBH (PZM02) and there are hints that 
a large fraction of the mass is lost due to winds and/or to recoil (Gaburov, Lombardi, Portegies Zwart 2010). 
Thus, there is no evidence that the MF of IMBHs is the same as that of the host YCs. Furthermore, even 
the density ($n_{\rm YC}(z,m_{\rm BH})$) and the MF of YCs are uncertain: in order to derive such quantities, 
we extrapolated to massive ($\ge{}10^4\,{}{\rm M}_\odot{}$) YCs various relationships  (e.g. the slope of the MF) derived by Lada \&{} Lada (2003) for a sample of 
smaller clusters ($20-1000\,{}{\rm M}_\odot{}$).

Finally, the models adopted to derive the instrumental range, and the  
corresponding $z_{\rm max}$, of the considered interferometers assume  
that the binary has zero eccentricity, the BHs are non-spinning and  
consider waveforms in a mass-ratio regime in which we still do not  
have reliable signal approximants. These are all factors that may  
significantly alter the results presented in this paper, in particular  
should YCs produce rapidly spinning and/or highly eccentric binaries.  
Full inspiral-merger-ringdown gravitational-waveform families for  
(generic) binary systems with a mass ratio of the order of 1$-$to$-$100  
are not available at present and we will be able to address this  
problem more rigorously only in the future.

We also note that, although we have consistently taken into account the
non-linear dynamics of the binary during the final merger by adopting full
inspiral-merger-ringdown waveforms (Ajith et al 2008a), calibrated on actual
numerical relativity simulations of the coalescence of BH binaries,
we have however neglected in the merger rate estimates the possibility that
the GW recoil (see, e.g. Baker et al. 2008; Lousto et al. 2010) may impart to the binary a sufficient recoil velocity to eject it from
the cluster. This would  prevent future mergers, and in this
respect, the results presented in this paper should be regarded as
upper-limits, as GW recoil can only reduce the merger rate.  On the other hand, assuming a typical escape velocity of 20 km s$^{-1}$ from the core of the host cluster (see e.g. Colpi et al. 2003), the GW recoil (calculated according to Baker et al. 2008) should not be able to eject IMBHs more massive than $\sim{}250$ M$_\odot{}$ ($\sim{}35$ M$_\odot{}$) merging with a 10 M$_\odot{}$ stellar mass BH (with a 1.4 M$_\odot{}$ NS). We notice, however, that the available expressions of the recoil velocity were obtained from simulations with mass ratio $m_{\rm co}/m_{\rm BH}=1-1/3$ (e.g Baker et al. 2008), whereas we consider systems for which $m_{\rm co}/m_{\rm BH}=0.1-0.0014$. No simulations have been carried out for such extreme mass ratio. Furthermore, even in a pessimistic scenario, accounting for GW recoils should reduce the detection rate $R$ by at most a factor of a few, as in our model most of IMBHs undergo $\lesssim{}5$ merger events within the lifetime of the cluster.

All these $caveats$ must be taken into account, 
when comparing the rates $R$ derived in this Paper with the forthcoming observational data.
Yet our results will provide the basis for future studies exploring these issues further
with improved data on the nature and frequency of YCs.

\acknowledgments
We thank the referee for his/her critical reading of the manuscript. 
MM acknowledges support from the Swiss
National Science Foundation, project number 200020-109581/1.


%% in parentheses, after the keyword, but they will not be verified.



%\clearpage

\appendix
\section{Method to estimate $z_{\rm max}$}
 We define $z_{\rm max}(m_{\rm BH},\,{}m_{\rm co})$ as the maximum redshift at which an event can be detected  with a sky location and orientation averaged
 signal-to-noise ratio $\langle \mathrm{SNR}\rangle\ge{}10$ by a %certain 
 single interferometer. The coherent SNR at which a gravitational wave $h(t)$ can be detected by an instrument characterized by a (zero-mean) Gaussian, stationary noise with a one-sided
 noise power spectral density $S_n(f)$, is given by 
\begin{equation}\label{eq:innprod}
{\rm SNR}^2= ({}h|h{}).
\end{equation}
Here, $({}.|.{})$ is the noise-weighted inner product, defined as (Cutler and Flanagan 1994)
\begin{equation}
({}a|b{})=2\int_0^\infty{}\frac{\tilde{a}^\ast{}(f)\,{}\tilde{b}(f)+\tilde{a}(f)\,{}\tilde{b}^\ast{}(f)}{S_n(f)}\,{}df\,,
\label{e:ab}
\end{equation}
where $\tilde{a}(f)$ and $\tilde{b}(f)$ are two generic functions defined in the Fourier domain.

For the  waveform $h(t)$ we use the phenomenological inspiral-merger-ringdown waveform model by Ajith et al. (2008a) for non-spinning BHs in circular orbit. For an {\it optimally orientated} binary the waveform emitted by during the whole coalescence is described by:
\begin{equation}
u(f)\equiv{}A_{\rm eff}(f)\,{}e^{\rm i \Psi{}_{\rm eff}(f)},
\end{equation}
where
\begin{equation}
A_{\rm eff}(f)\equiv{}C\left\{
\begin{array}{l}
(f/f_{\rm merg})^{-7/6} \textrm{  if } f<f_{\rm merg}\\
(f/f_{\rm merg})^{-2/3} \textrm{  if } f_{\rm merg}\le{} f < f_{\rm ring}\\
w\mathcal{L}(f,f_{\rm ring},\sigma{}) \textrm{  if } f_{\rm ring}\le{}f<f_{\rm cut}
\end{array}
\right.
\label{e:uGW}
\end{equation}
Expressions for $f_{\rm merg}$, $f_{\rm ring}$, $f_{\rm cut}$, $\Psi{}_{\rm eff}(f)$, $C$, $w$ and  $\mathcal{L}(f,f_{\rm ring},\sigma{})$ are given in equations (4.14-4.19) and Tables I-II of Ajith et al. (2008a). The fitting coefficients reported in Ajith et al. (2008a) have been revised in Ajith (2008b), but have no effect on the actual signal-to-noise ratio, and we have therefore adopted the original values. The angle-averaged signal-to-noise ratio is obtained by dividing the SNR, equation~(\ref{eq:innprod}) from an optimally oriented source, described by equation~(\ref{e:uGW}), by a factor 2.26, which accounts for the varying response of GW instruments to sources in different locations of the sky and with different orientations of the orbital angular momentum:
\begin{equation}
\langle \mathrm{SNR}\rangle = \frac{({}u|u{})}{2.26}
\label{eq:snrav}
\end{equation}
In observations carried out with a network the total coherent network SNR is $\langle \mathrm{SNR}\rangle^2 = \sum_k \langle \mathrm{SNR}\rangle^2_k$, where $\langle \mathrm{SNR}\rangle^2_k$ is the signal-to-noise ratio at each instrument. We caution the
reader that if sources are not uniformly distributed in distance, then
considering angle-averaged signal-to-noise ratios (equation~(\ref{eq:snrav})) 
introduces some
errors in the estimate of the detection rates; however, considering the
other (large) uncertainties (from the astrophysics and waveform modelling)
that enter the computation of this quantity, the simplification introduced
by equation~(\ref{eq:snrav}) has a negligible effect on the final results.

Using these equations, we can derive the luminosity distance $D_L(z_{\rm max}(m_{\rm BH},\,{}m_{\rm co}))$ at which an event can be detected  with a ${\rm SNR}\ge{}10$ by a certain interferometer. 
We can, thus, derive $z_{\rm max}(m_{\rm BH},\,{}m_{\rm co})$ by inverting the expression of the luminosity distance in the $\Lambda{}$ Cold Dark Matter (CDM) model:
\begin{equation}
D_L(z_{\rm max}(m_{\rm BH},\,{}m_{\rm co}))=\frac{c}{H_0}(1+z_{\rm max}(m_{\rm BH},\,{}m_{\rm co}))\,{}\int_0^{z_{\rm max}{(m_{\rm BH},\,{}m_{\rm co})}}{\frac{{\rm d}z}{E(z)}}
\label{e:DL}
\end{equation}

The range of an instrument, equation~(\ref{e:DL}) and therefore the IMBH binary detection rates are entirely determined by the noise spectral density $S_n(f)$, see equation~(\ref{eq:innprod}). For ground-based interferometers we consider representative sensitivity curves for the three generations of instruments. For instruments now in operation (first generation) we adopt the initial LIGO design sensitivity curve, that well approximates the sensitivity achieved during the last science run (Abbott et al. 2009). Consistently we set the low frequency cut-off, the minimum of integration in equation~(\ref{e:ab}), to $f_\mathrm{min} = 40\,\mathrm{Hz}$; the upper frequency cut-off is irrelevant, as the ring-down signal for IMBH mass-range of interest is \emph{de facto} zero in the high-frequency region of the instrumental sensitivity window $f\sim 1$ kHz.

 For second-generation (or advanced) interferometers we use the broad band target design sensitivity curve of Advanced LIGO with a low frequency cut-off $f_\mathrm{min} = 10\,\mathrm{Hz}$ (Adhikari et al, 2006); Advanced Virgo is expected to operate on the same timescale and has similar noise performance. 

For third generation interferometers, that are currently undergoing conceptual design studies, we adopt the noise curve of a single right-angle ET instrument and a low frequency cut-off $f_\mathrm{min} = 1$ Hz (Hild et al. 2008). We note that other configurations, such as the Xylophone (Hild et al. 2010), have been proposed and are currently under study.

For LISA we adopt the current best estimate of the \emph{instrumental} noise spectral density, see Barack and Cutler (2004). As we show below, the frequency range relevant for observations of IMBH binaries is always above several mHz; as a consequence, the confusion noise generated by stochastic foregrounds of close white-dwarf binaries (and possibly extreme mass-ratio inspirals) does not contribute to the total noise budget and we therefore ignore it in the calculation. In the case of low-frequency observations, the binary system lifetime may be longer than the mission lifetime. In fact, the time to coalescence (at the leading quadrupole Newtonian order) for a binary radiating at frequency $f$ is
\begin{equation}
\tau(f) = 5\,{}c^5\,{}G^{-5/3}\,{}(8\,{}\pi{}\,{}f)^{-8/3}\,{} \mu{}^{-1}\,{}m_{\rm tot}^{-2/3}\,,
\end{equation}
where $G$ is the gravitational constant, $c$ is the speed of light, and $m_{\rm tot}=m_{\rm BH}+m_{\rm co}$ and $\mu{}=m_{\rm BH}\,{}m_{\rm co}/m_{\rm tot}$ are the total and reduced mass, respectively, of the coalescing binary. 
Moreover, LISA will not be able to observe the final merger-ringdown phase, as it takes place at frequency higher than the observable window. Here we assume that the highest frequency that LISA can observe is 1 Hz and the last stable orbit around a Schwarzschild BH corresponds to the GW frequency $\approx 4\,{}(10^3\,{\rm M}_\odot/m_{\rm tot})$ Hz. The integration limits in equation~(\ref{eq:innprod}) are obtained with the following procedure. The time of observability ${\rm T}_{\rm obs}$ for a merger event is given by (Peters 1964)
\begin{equation}\label{eq:tobs}
{\rm T}_{\rm obs}= \tau(f_\mathrm{min}) -  \tau(f_\mathrm{max})\,.
\end{equation}
From equation~(\ref{eq:tobs}), we can derive $f_{\rm min}$ as a function of $f_{\rm max}$, assuming ${\rm T}_{\rm obs}=5$ yr.
 Fig.~\ref{fig:fig2} shows $f_{\rm min}$ as a function of $f_{\rm max}$. We note that $f_{\rm min}$ is almost constant for $f_{\rm max}\gtrsim{}3\times{}10^{-2}$ Hz.  We then calculate $\langle{}{\rm SNR}\rangle{}$ from equation~(\ref{eq:snrav}) for different values of the couple $f_{\rm min}$, $f_{\rm max}$. Fig.~\ref{fig:fig3} shows $\langle{}{\rm SNR}\rangle{}$ as a function of the couple $f_{\rm min}$, $f_{\rm max}$. We find that  $\langle{}{\rm SNR}\rangle{}$  is maximum when we take $f_{\rm max}=1$ Hz and we consistently derive $f_{\rm min}$ from equation~(\ref{eq:tobs}). The results presented in Table~1 and in Fig.~\ref{fig:fig1} are obtained with such choice of $f_{\rm max}=1$ Hz and $f_{\rm min}$ derived from equation~(\ref{eq:tobs}). As a consequence the results presented for LISA in this Paper should be considered as upper limits, assuming that the binary is in the optimal stage of the merger for detection. However, from Fig.~\ref{fig:fig3} we also note that $\langle{}{\rm SNR}\rangle{}$ is almost constant for $f_{\rm max}\ge{}10^{-2}$ Hz. Thus, we expect that the upper limit for LISA is not far from the average value. The integration frequency range also justifies neglecting the confusion noise from unresolved stochastic foregrounds.

\section{Alternative derivation of $R$}
Gair et al. (2009b) report in their equation~(14) a derivation of the detectable merger rate for IMBH$-$IMBH binaries. In the current paper, we do not consider the mergers of IMBH$-$IMBH binaries, because the possibility of forming more than one IMBH in the same cluster is still debated (see e.g. G\"urkan, Fregeau \&{} Rasio 2006). However, it is possible to perform a calculation similar to equation~(14) of Gair et al. (2009b) for IMBH$-$BH and IMBH$-$NS binaries, as follows.
\begin{equation}\label{eq:gair}
R=\int_{m_1}^{m_2}{\rm d}m_{\rm BH}\int_0^{z_{\rm max}(m_{\rm BH}, m_{\rm co})}{\rm d}\tilde{z}\frac{{\rm d}^3N_{\rm merg}}{{\rm d}m_{\rm BH}\,{}{\rm d}t_{\rm e}\,{}{\rm d}V_{\rm c}}\,{}\frac{{\rm d}t_{\rm e}}{{\rm d}{t_{\rm o}}}\,{}\frac{{\rm d}V_{\rm c}}{{\rm d}\tilde{z}},
\end{equation}
where $m_1$ and $m_2$ are the minimum and the maximum IMBH mass, respectively;  $\frac{{\rm d}t_{\rm e}}{{\rm d}{t_{\rm o}}}=(1+z)^{-1}$ (since $t_{\rm e}$ and $t_{\rm o}$ are the time in the rest frame of the source and of the observer, respectively); $V_{\rm c}$ is the comoving volume; $N_{\rm merg}$ is the number of mergers. Assuming that $m_{\rm BH}=10^{-3}m_{\rm YC}$ (PZM02),
\begin{equation}\label{eq:gair2}
\frac{{\rm d}^3N_{\rm merg}}{{\rm d}m_{\rm BH}\,{}{\rm d}t_{\rm e}\,{}{\rm d}V_{\rm c}}=\nu{}_{\rm mrg}(m_{\rm BH})\,{}f_{\rm tot}\,{}t_{\rm max}\,{}f_{\rm surv}\,{}10^3\frac{{\rm d}^3N_{\rm YC}}{{\rm d}m_{\rm YC}\,{}{\rm d}t_{\rm e}\,{}{\rm d}V_{\rm c}},
\end{equation}
where  $N_{\rm YC}$ is the number of YCs. Following equation~(12) of Gair et al. (2009b)
\begin{equation}\label{eq:gair3}
\frac{{\rm d}^3N_{\rm YC}}{{\rm d}m_{\rm YC}\,{}{\rm d}t_{\rm e}\,{}{\rm d}V_{\rm c}}=\frac{g_{\rm YC}(m_{\rm YC})}{\ln{(m_{\rm YC,max}/m_{\rm YC,min})}}\,{}\dot{\rho{}}_{\ast{}}(z)\,{}m_{\rm YC}^{-2},
\end{equation}
where $\dot{\rho{}}_{\ast{}}(z)$ is the comoving density of SFR (see Section 2.3) and $g_{\rm YC}(m_{\rm YC})$ is the fraction of the total stellar mass which is formed in YCs of mass $m_{\rm YC}$.  

Substituting equation~(\ref{eq:gair3}) into equation~(\ref{eq:gair}), and assuming that $\nu{}_{\rm mrg}(m_{\rm BH})=10^{-2}\,{}\nu{}_{3b}(m_{\rm BH})$ (see Section 2.1) and $m_{\rm BH}=10^{-3}m_{\rm YC}$, we finally obtain:
\begin{eqnarray}\label{eq:gair4}
R=\frac{2\,{}\pi{}\,{}G\,{}f_{\rm tot}\,{}t_{\rm max}\,{}f_{\rm surv}}{\ln{(m_{\rm YC,max}/m_{\rm YC,min})}}\,{}10^{-5}\,{}n_{\rm c}\,{}a\,{}\sigma{}_{\rm c}^{-1}\,{}g_{\rm YC}\,{}\int_{m_{\rm YC,minBH}}^{m_{\rm YC,maxBH}}\frac{{\rm d}m_{\rm YC}}{m_{\rm YC}}\nonumber{}\\\,{}
\times{}\int_0^{z_{\rm max}(10^{-3}\,{}m_{\rm YC}, m_{\rm co})}{\rm d}\tilde{z}\,{}
\frac{\dot{\rho{}}_{\ast{}}(\tilde{z})}{(1+\tilde{z})}\frac{{\rm d}V_{\rm c}}{{\rm d}\tilde{z}},
\end{eqnarray}
where $m_{\rm YC, minBH}$ and $m_{\rm YC, maxBH}$ are the minimum and the maximum YC mass in order to form an IMBH in the considered mass range, respectively (we adopt $m_{\rm YC,minBH}=10^5$ M$_\odot{}$ and $m_{\rm YC, maxBH}=10^6$ M$_\odot{}$, corresponding to $m_{\rm BH}$ between 10$^2$ and 10$^3$ M$_\odot{}$). As in Section~2.3, $m_{\rm YC,min}=20$ M$_\odot{}$ and $m_{\rm YC,max}=10^7$ M$_\odot{}$ are the minimum and the maximum YC mass. In equation~(\ref{eq:gair4}) we assume that $n_{\rm c}$, $a$, $\sigma{}_{\rm c}$ and $g_{\rm YC}$ do not depend on the cluster mass. Adopting  $n_{\rm c}=5\times{}10^5$ pc$^{-3}$, $a=0.4$ AU, $\sigma{}_{\rm c}=20$ km s$^{-1}$ and $g_{\rm YC}=0.8$ (the same values as in the alternative calculation reported in the main text), we derive the values of $R$ shown in Fig.~\ref{fig:fig1} as open points (open triangles for LISA, open circles for Advanced LIGO and open squares for ET). The difference between open and filled points (derived from equation~\ref{eq:gair4} and from equation~\ref{eq:GW2}, respectively) is within the large uncertainties of the two calculations. In Table~2, we report the values of $R$ derived from equation~(\ref{eq:gair4}), for a comparison with those listed in Table~1: differences are less than a factor of 7 in the case of LISA and less than a factor of 2 for the other instruments.

\section{Approximate derivation of $R$}
\begin{figure}
\epsscale{.80}
\plotone{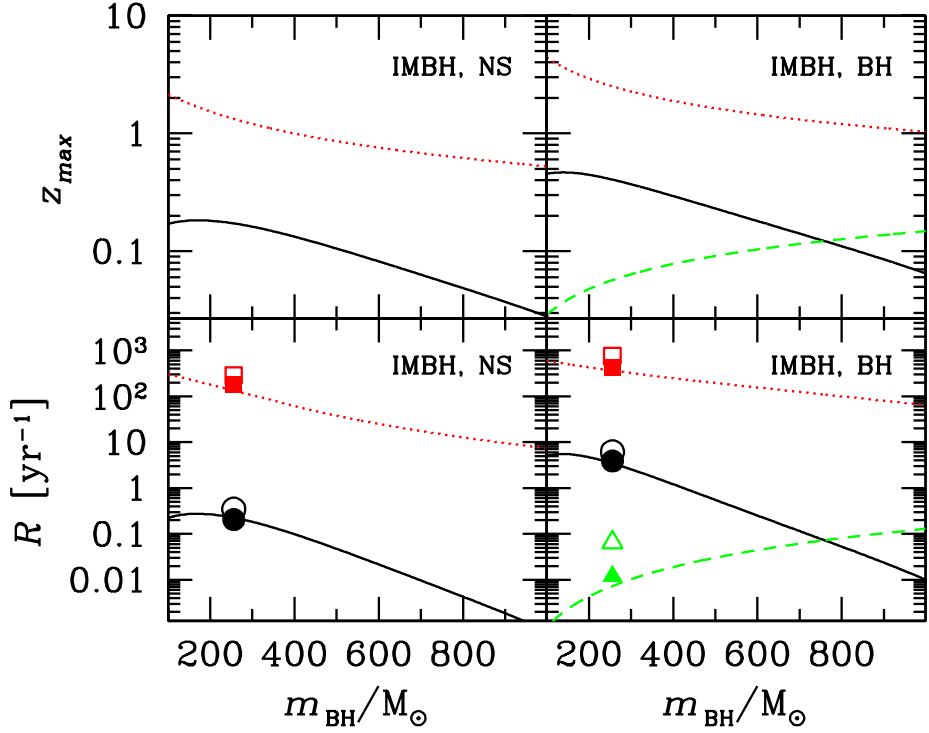}
\caption{Top panels: $z_{\rm max}(m_{\rm BH},\,{}m_{\rm co})$ as a function of the  IMBH mass $m_{\rm BH}$ for IMBHs merging with NSs (left-hand panel) and with stellar mass BHs (right-hand panels). Bottom panels: detection rate $R$ as a function of the IMBH mass $m_{\rm BH}$ for IMBHs merging with NSs (left-hand panel) and with stellar mass BHs (right-hand panels). 
For all the panels, dotted line (red on the web): events detectable by ET; solid black line: events detectable by Advanced LIGO; dashed line (green on the web): events detectable by LISA. 
The events detectable by LIGO are not shown  in this Fig., because they are more than one order of magnitude below the limits of the $y-$axes. For the same reason, $z_{\rm max}(m_{\rm BH},\,{}m_{\rm co})$ and $R$  for IMBH$-$NS mergers are not shown in the case of LISA.
In the bottom panels, the filled square (red on the web), the black filled circle and the filled triangle (green on the web) are the detection rate $R$ in the case of ET, Advanced LIGO and LISA, respectively, obtained from equation~(\ref{eq:GW2}), assuming a Lada \&{} Lada (2003) MF for the IMBHs. These points are drawn, for simplicity, in correspondence of $m_{\rm BH}=256\,{}{\rm M}_\odot$, but they have been obtained integrating equation~(\ref{eq:GW2}) over a Lada \&{} Lada (2003) MF, with an average IMBH mass $\langle{}m_{\rm BH}\rangle{}=256\,{}{\rm M}_\odot{}$ (see Section 2.3 for details). In the bottom panels, the open square (red on the web), the black open circle and the open triangle (green on the web) are the detection rate $R$ in the case of ET, Advanced LIGO and LISA, respectively, obtained from equation~(\ref{eq:gair4}), assuming a Lada \&{} Lada (2003) MF for the IMBHs. 
\label{fig:fig1}}
\end{figure}

\begin{figure}
\epsscale{.80}
\plotone{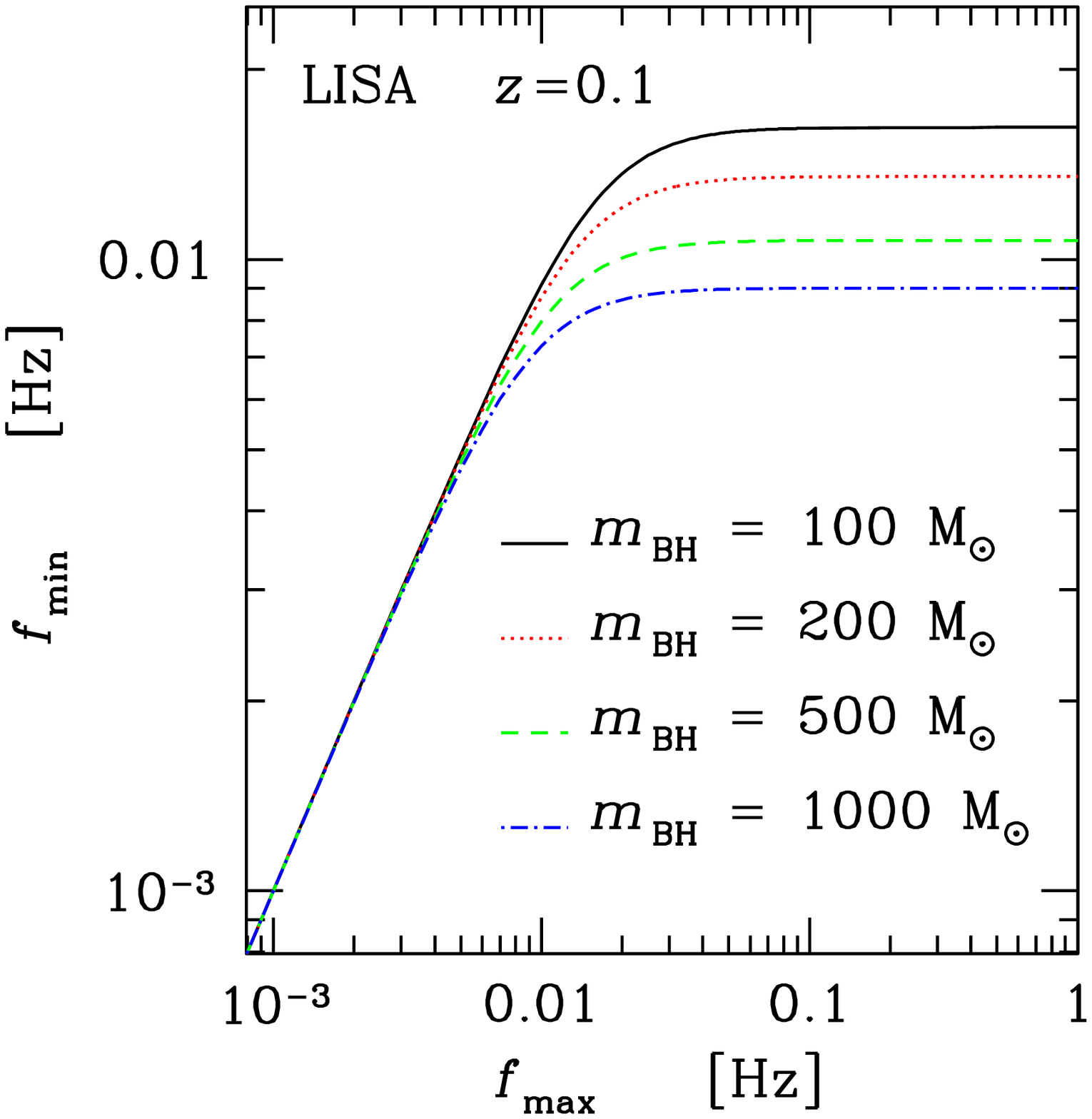}
\caption{
$f_{\rm min}$ as a function of $f_{\rm max}$, in the case of LISA, for a binary at $z=0.1$ with $m_{\rm co}=10\,{}{\rm M}_\odot{}$. Solid black line: $m_{\rm BH}=100\,{}{\rm M}_\odot{}$; dotted line (red on the web): $m_{\rm BH}=200\,{}{\rm M}_\odot{}$; dashed line (green on the web): $m_{\rm BH}=500\,{}{\rm M}_\odot{}$; dot-dashed line (blue on the web): $m_{\rm BH}=1000\,{}{\rm M}_\odot{}$. 
\label{fig:fig2}}
\end{figure}

\begin{figure}
\epsscale{.80}
\plotone{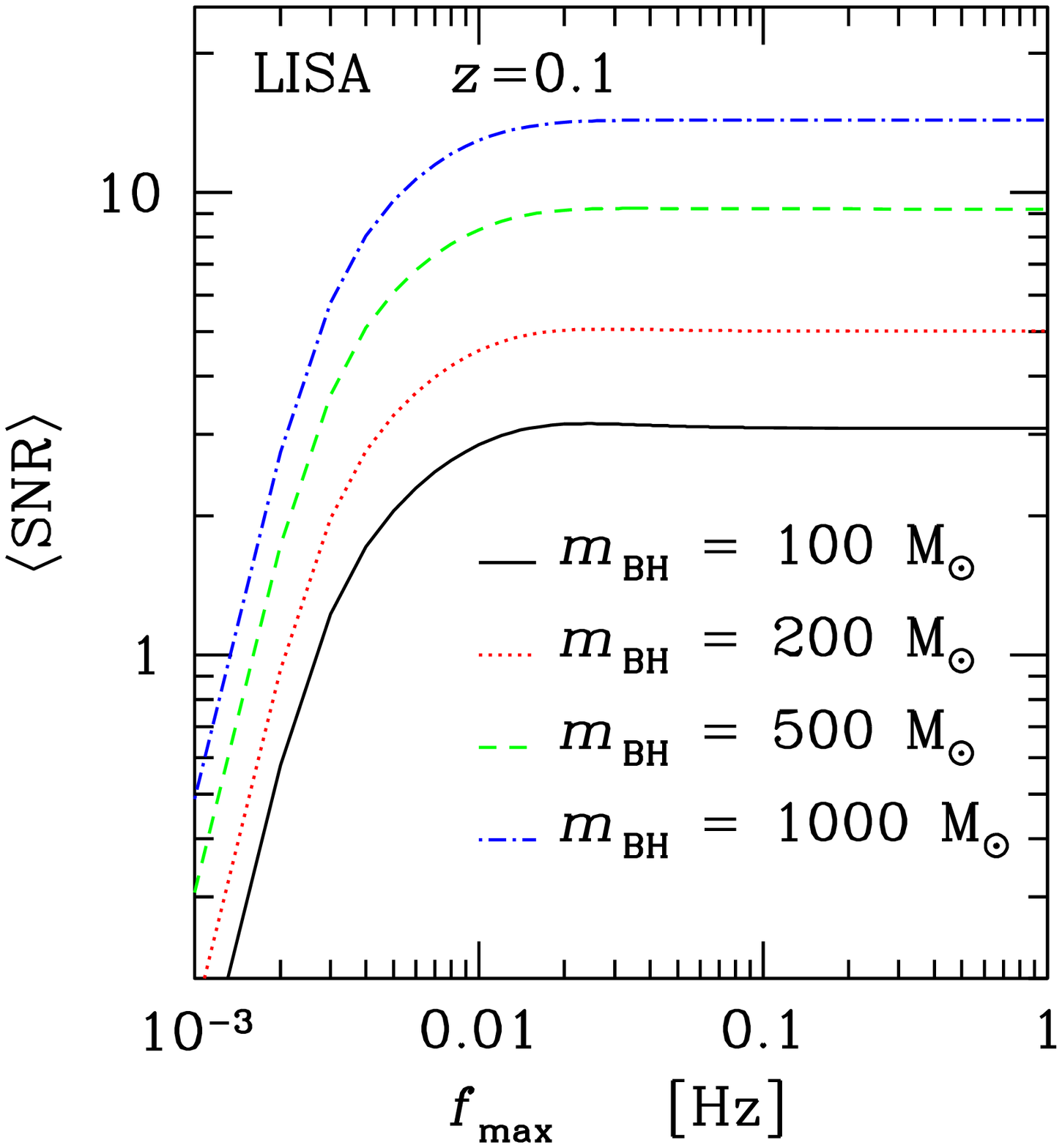}
\caption{
$\langle{}{\rm SNR}\rangle{}$ as a function of $f_{\rm max}$, in the case of LISA, for a binary at $z=0.1$ with $m_{\rm co}=10\,{}{\rm M}_\odot{}$. Solid black line: $m_{\rm BH}=100\,{}{\rm M}_\odot{}$; dotted line (red on the web): $m_{\rm BH}=200\,{}{\rm M}_\odot{}$; dashed line (green on the web): $m_{\rm BH}=500\,{}{\rm M}_\odot{}$; dot-dashed line (blue on the web): $m_{\rm BH}=1000\,{}{\rm M}_\odot{}$. 
\label{fig:fig3}}
\end{figure}

\begin{figure}
\epsscale{.80}
\plotone{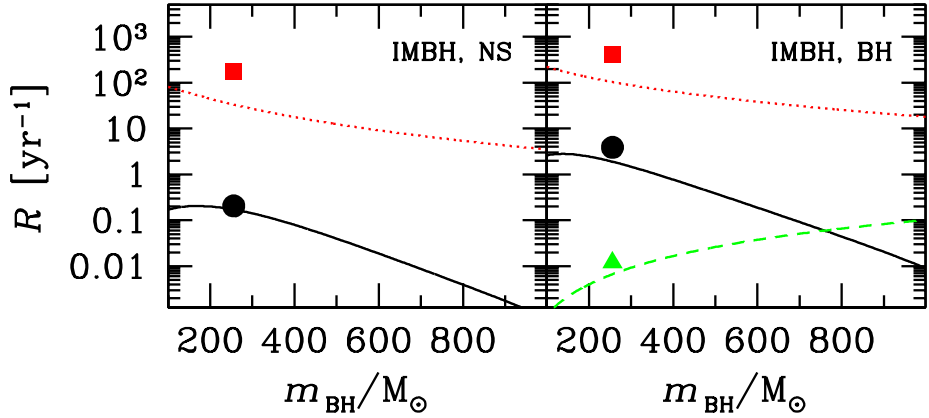}
\caption{
Left-hand (Right-hand) panel: detection rate $R$ for IMBH$-$NS (IMBH$-$BH) mergers derived with the order-of-magnitude calculation reported in equation~(\ref{eq:approx}).  Dotted line (red on the web): events detectable by ET; solid black line: events detectable by Advanced LIGO; dashed line (green on the web): events detectable by LISA.  For comparison, we include also the results of equation~(\ref{eq:GW2}). In particular, the filled square (red on the web), the black filled circle and the filled triangle (green on the web) are the same as in Fig.~\ref{fig:fig1}, i.e. the detection rate $R$ in the case of ET, Advanced LIGO and LISA, respectively, obtained from equation~(\ref{eq:GW2}), assuming a Lada \&{} Lada (2003) MF for the IMBHs.
\label{fig:fig4}}
\end{figure}

An order-of-magnitude estimate of $R$ may be obtained with a much simpler calculation than those reported in Section 2.3 and in Appendix~B. In particular, assuming that all the IMBHs have the same mass $m_{\rm BH}$, we can write: 
\begin{equation}\label{eq:approx}
R\approx{}f_{\rm tot}\,{}\frac{n_{\rm YC}(z=0, m_{\rm BH})}{\rm T_{\rm mrg}(m_{\rm BH})}\,{}V_{\rm c}(z_{\rm max}), 
\end{equation}
where $n_{\rm YC}(z=0, m_{\rm BH})$ is the comoving density of YCs which are sufficiently massive to host an IMBH, defined by the equation~(\ref{eq:eqYC}), calculated at $z=0$. We note that we are considering a lower limit of the YC density, as the value of $n_{\rm YC}(z, m_{\rm BH})$ is minimum for $z=0$. T$_{\rm mrg}=\nu{}_{\rm mrg}^{-1}=2\times{}10^8$ yr $(256\,{}{\rm M}_\odot{}/m_{\rm BH})$ $(5\times{}10^5{\rm pc}^{-3}/n_{\rm c})$ $(\sigma{}_{\rm c}/20\,{}{\rm km s}^{-1})$ is the  timescale for a merger. Finally, 
$V_{\rm c}(z_{\rm max})$ is the comoving volume up to redshift $z_{\rm max}(m_{\rm BH})$ (with $z_{\rm max}(m_{\rm BH})$ derived as described in Appendix~A). The results of equation~(\ref{eq:approx}) as a function of the fiducial IMBH mass are shown in Fig.~\ref{fig:fig4}. Table~3 reports the value of $R$ for a fixed IMBH mass $m_{\rm BH}=256$ M$_\odot{}$. We note that the detection rates predicted by equation~(\ref{eq:approx}) are a factor of $\sim{}2-10$ lower than those derived with the other two, more refined, methods (see Fig.~\ref{fig:fig1}). This is likely due to the fact that in equation~(\ref{eq:approx}) $n_{\rm YC}$ does not increase with redshift. For this reason, equation~(\ref{eq:approx}) is a sort of lower limit for our estimates.

Finally, Table~4 is focused on the case of ET, and  shows both the quantities adopted in equation~(\ref{eq:approx}) and the derived detection rates for some choices of $m_{\rm BH}$ and $m_{\rm co}$. Table~4 was introduced for a comparison between our procedure and the similar one reported in Section 3.1 and in Table~1 of Gair et al. (2009b). We note that the values of the comoving density of young clusters which are sufficiently massive to host an IMBH ($n_{\rm YC}$), adopted in our calculations, are a factor of $3-40$ lower than those adopted by Gair et al. (2009b), who consider the comoving space density of globular clusters and assume that $\sim{}10$ per cent of them host an IMBH. On the other hand, the usage of Ajith et at. (2008a) model for the gravitational waveform leads to an overestimate of $z_{\rm max}$ by a factor of $2-3$ with respect to the EOBNR models adopted by Gair et al. (2009b, see the discussion in the main text for details). For these reasons, our estimates of $R$ are very similar (within a factor of two) to those reported in Table~1 of Gair et al. (2009b).

%%%%%%%%%%%%%%%%%%%%%%%%%%%%%%%%FIGURE 1%%%%%%%%%%%%%%%%%%%%%%%%%%%%%%%%%%%%%%%

%%%%%%%%%%%%%%%%%%%%%%%%%%%%%%%%%%%%%%%%%%%%%%%%%%%%%%%%%%%%%%%%%%%%%%%%%%%%%%%
\clearpage
%%%%%%%%%%%%%%%%%%%%%%%%%%%%%%%%FIGURE 2%%%%%%%%%%%%%%%%%%%%%%%%%%%%%%%%%%%%%%%

%%%%%%%%%%%%%%%%%%%%%%%%%%%%%%%%%%%%%%%%%%%%%%%%%%%%%%%%%%%%%%%%%%%%%%%%%%%%%%%

\clearpage

%%%%%%%%%%%%%%%%%%%%%%%%%%%%%%%%FIGURE 3%%%%%%%%%%%%%%%%%%%%%%%%%%%%%%%%%%%%%%%

%%%%%%%%%%%%%%%%%%%%%%%%%%%%%%%%%%%%%%%%%%%%%%%%%%%%%%%%%%%%%%%%%%%%%%%%%%%%%%%

\clearpage
%%%%%%%%%%%%%%%%%%%%%%%%%%%%%%%%FIGURE 4%%%%%%%%%%%%%%%%%%%%%%%%%%%%%%%%%%%%%%%

%%%%%%%%%%%%%%%%%%%%%%%%%%%%%%%%%%%%%%%%%%%%%%%%%%%%%%%%%%%%%%%%%%%%%%%%%%%%%%%

\clearpage
%%%%%%%%%%%%%%%%%%%%%%%%%%%%%%%%%%TABLE 1%%%%%%%%%%%%%%%%%%%%%%%%%%%%%%%%%%%%%%
\begin{table}
\begin{center}
\caption{{Detection rates from equation~(\ref{eq:GW2}) assuming a Lada \&{} Lada (2003) MF and $f_{\rm tot}=0.75$ for the IMBHs (see Section 2.1 and 2.3 for details).}} 
\footnotesize
\begin{tabular}{l|c|c|c|c}
\tableline\tableline
& {\bf{LIGO}} & {\bf{Advanced LIGO}} & \bf{LISA} & \bf{ET} \\
& $R$ [yr$^{-1}$] & $R$ [yr$^{-1}$] & $R$ [yr$^{-1}$] & $R$ [yr$^{-1}$]\\
\tableline
\bf{NS}          & $2\times{}10^{-5}$ & $0.2$ & $2\times{}10^{-5}$  & $200$ \\
\bf{stellar mass BHs} & $3\times{}10^{-4}$ & $4$   & $0.01$              & $400$ \\
\tableline
\end{tabular}
\footnotesize
\end{center}
\end{table}
%%%%%%%%%%%%%%%%%%%%%%%%%%%%%%%%%%%%%%%%%%%%%%%%%%%%%%%%%%%%%%%%%%%%%%%%%%%%%%%

\clearpage
%%%%%%%%%%%%%%%%%%%%%%%%%%%%%%%%%%TABLE B1%%%%%%%%%%%%%%%%%%%%%%%%%%%%%%%%%%%%%
\begin{table}
\begin{center}
\caption{{Detection rates from equation~(\ref{eq:gair4}) assuming a Lada \&{} Lada (2003) MF  and $f_{\rm tot}=0.75$ for the IMBHs (see Appendix~B for details).}} 
\footnotesize
\begin{tabular}{l|c|c|c|c}
\tableline\tableline
& {\bf{LIGO}} & {\bf{Advanced LIGO}} & \bf{LISA} & \bf{ET} \\
& $R$ [yr$^{-1}$] & $R$ [yr$^{-1}$] & $R$ [yr$^{-1}$] & $R$ [yr$^{-1}$]\\
\tableline
\bf{NS}          & $2\times{}10^{-5}$ & $0.3$ & $1\times{}10^{-4}$  & $300$ \\
\bf{stellar mass BHs} & $3\times{}10^{-4}$ & $6$   & $0.07$              & $750$ \\
\tableline
\end{tabular}
\footnotesize
\end{center}
\end{table}
%%%%%%%%%%%%%%%%%%%%%%%%%%%%%%%%%%%%%%%%%%%%%%%%%%%%%%%%%%%%%%%%%%%%%%%%%%%%%%%

\clearpage
%%%%%%%%%%%%%%%%%%%%%%%%%%%%%%%%%%TABLE C1%%%%%%%%%%%%%%%%%%%%%%%%%%%%%%%%%%%%%
\begin{table}
\begin{center}
\caption{{Detection rates from equation~(\ref{eq:approx}) assuming  $m_{\rm BH}=256$ M$_\odot{}$ and $f_{\rm tot}=0.75$ for the IMBHs (see Appendix~C for details).}} 
\footnotesize
\begin{tabular}{l|c|c|c|c}
\tableline\tableline
& {\bf{LIGO}} & {\bf{Advanced LIGO}} & \bf{LISA} & \bf{ET} \\
& $R$ [yr$^{-1}$] & $R$ [yr$^{-1}$] & $R$ [yr$^{-1}$] & $R$ [yr$^{-1}$]\\
\tableline
\bf{NS}          & $2\times{}10^{-6}$ & $0.2$ & $1\times{}10^{-5}$  & $30$ \\
\bf{stellar mass BHs} & $3\times{}10^{-5}$ & $2$   & $7\times{}10^{-3}$  & $100$ \\
\tableline
\end{tabular}
\footnotesize
\end{center}
\end{table}
%%%%%%%%%%%%%%%%%%%%%%%%%%%%%%%%%%%%%%%%%%%%%%%%%%%%%%%%%%%%%%%%%%%%%%%%%%%%%%%

\clearpage
%%%%%%%%%%%%%%%%%%%%%%%%%%%%%%%%%%TABLE C2%%%%%%%%%%%%%%%%%%%%%%%%%%%%%%%%%%%%%
\begin{table}
\begin{center}
\caption{{Quantities used in  equation~(\ref{eq:approx}) and detection rates $R$, derived from the same equation, in the case of ET (see Appendix~C for details).}} 
\footnotesize
\begin{tabular}{ccccccc}
\tableline\tableline
$m_{\rm BH}$ [M$_\odot{}$] & $m_{\rm co}$  [M$_\odot{}$] & $z_{\rm max}$ & n$_{\rm YC}$ [Mpc$^{-3}$] & T$_{\rm mrg}$ [yr] & V$_{\rm c}$ [Mpc$^3$] & $R$ [yr$^{-1}$] \\
\tableline
100 & 1.4 & 2.1 & 0.09   & $5.2\times{}10^8$ & $6.1\times{}10^{11}$ & 80\\
100 & 10  & 4.4 & 0.09   & $5.2\times{}10^8$ & $1.7\times{}10^{12}$ & 200\\
300 & 1.4 & 1.2 & 0.03   & $1.7\times{}10^8$ & $2.1\times{}10^{11}$ & 30\\
300 & 10  & 2.3 & 0.03   & $1.7\times{}10^8$ & $6.9\times{}10^{11}$ & 90\\
1000 & 1.4 & 0.5 & 0.008 & $5.2\times{}10^7$ & $2.9\times{}10^{10}$ & 3\\
1000 & 10  & 1.0 & 0.008 & $5.2\times{}10^7$ & $1.5\times{}10^{11}$ & 20\\
\tableline
\end{tabular}
\footnotesize
\end{center}
\end{table}
%%%%%%%%%%%%%%%%%%%%%%%%%%%%%%%%%%%%%%%%%%%%%%%%%%%%%%%%%%%%%%%%%%%%%%%%%%%%%%%

\end{document}